# From Method Fragments to Method Services

Rébecca Deneckère, Adrian Iacovelli, Elena Kornyshova, Carine Souveyet

CRI, University Paris 1 – Panthéon Sorbonne, 90, rue de Tolbiac,
75013 Paris, France
{denecker,adrian.iacovelli,kornyshova,souveyet}@univ-paris1.fr

**Abstract.** In Method Engineering (ME) science, the key issue is the consideration of information system development methods as fragments. Numerous ME approaches have produced several definitions of method parts. Different in nature, these fragments have nevertheless some common disadvantages: lack of implementation tools, insufficient standardization effort, and so on. On the whole, the observed drawbacks are related to the shortage of usage orientation. We have proceeded to an in-depth analysis of existing method fragments within a comparison framework in order to identify their drawbacks. We suggest overcoming them by an improvement of the "method service" concept. In this paper, the method service is defined through the service paradigm applied to a specific method fragment – chunk. A discussion on the possibility to develop a unique representation of method fragment completes our contribution.

## 1 Introduction

Method engineering (ME) science deals with information systems (IS) development methods. One of the ME fundamentals for optimizing, reusing, and ensuring flexibility and adaptability of these methods is their decomposition into modular parts [1].

There are various representations of building blocks. This purpose is discussed in the literature [2, 3, 4] and gives an overview of five different building blocks: fragments [5], chunks [6], components [7], OPF fragments [8], and method services [9]. We will use the term "Fragment" in this work as a generic term for all kinds of building blocks. Historically, the term fragment was the first one to appear, long before component, chunk, and so on. Brinkkemper defines a method fragment as "a coherent piece of an IS development method" [5]. Therefore, we consider this definition as the simpler one and that all others are essentially its extensions, which is the reason why we have chosen this term. The description of fragments is strongly linked to the approaches that suggest them. For this reason, we consider the fragment definitions as joint notions of ME approaches.

Despite their diversity, different method fragments have some common drawbacks. To identify them, we elaborate a comparison framework. From the application of our comparison framework on the five selected fragments, we deduce that a sufficient tool support is not provided for them and for their use (interactivity with users). Moreover, the interoperability of the proposed proprietary solutions is not handled. In addition, the complexity of data exchanged is not completely addressed.

In order to overcome these drawbacks, we suggest improving Rolland's proposal [10] about applying the service-based approach to ME needs. This concerns the adaptation of Service Oriented Architecture (SOA) [11] to method fragment by developing a Method Oriented Architecture (MOA). In this manner, we improve the concept of "method service".

Our method service contains two parts: descriptor and implementation parts. The *descriptor part* combines a semantic descriptor (based on the fragment definition of the method chunks approach) and an operational descriptor describing the *implementation part* that operates the process of the fragment. Technical issues of method services are addressed with the application of widely used standards of service-based approaches.

Thus, this study joins the ME field with the proposal of a framework used for comparing different representations of method building blocks, for identifying their drawbacks, and suggesting a solution to solve them.

This paper is organised as follows. Our comparison framework is described in the next section and it is applied on three selected method fragments in the third section. Following the concluding remarks of this comparison, the concept of the method service is developed in section 4. A discussion about a unique concept of method fragments is addressed in section 5 and section 6 concludes this work with our contribution and research perspectives.

## 2 Comparison Framework

We have elaborated a framework to compare different method fragments. The idea to consider a central concept (here the method fragment) on four different points of view is largely inspired from [12], a work dealing with evolution scenarios. To elaborate our comparison framework, we have proceeded to an analysis of issues that are crucial for a "good" IS development method and, at the same time, not-solved by existing method Fragments. As a result, our framework contains 15 attributes organized into 4 views (cf. Fig. 1) developed in the following subsections.

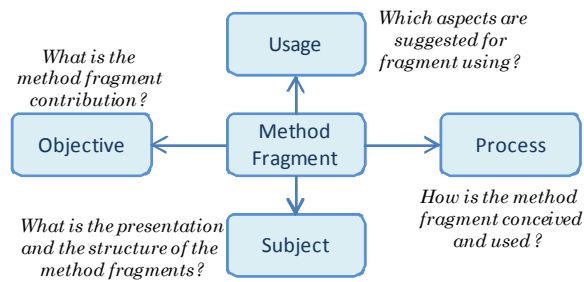

**Fig. 1.** Method fragments' comparison framework.

### 2.1 Objective View

This view captures why we should use a specific method fragment and what are the benefits retrieved from its practical application.

A point to consider is the fragment *interoperability* with other fragments. The interoperability has been discussed since the beginning of ME science [5]. However, the majority of fragments are conceived to be interoperable only with the fragments stored in the same method base ("internal" interoperability). In the real world project, the situation is widespread when the interoperability is required with other elements (external to method base) on the same or on different development platforms (external interoperability within or not the same environment).

Staying within the fragments' environment, benefits are retrieved from the degree of interactions with the method engineer. This *interactivity* is decomposed into three possible levels. ME approaches should either provide a fully automated or assisted (semi-automated) process for construction, reuse and composition of fragments. At least, the ME approach application can be manual.

### 2.2 Usage View

This view deals with different aspects that describe the fragment usage.

Seligmann gives a definition of a method as "a way of thinking, a way of modelling, a way of working and a way of supporting" [13]. However, even if a lot of fragments are considered as complete method, often they are not adapted to satisfy all these requirements. We investigate this question through the *covered way*.

The methods fragments application needs to be supported by a tool. [5] defines a tool as "possibly automated means to support a parts of development process". We distinguish different ways of fragment implementation: first, the implementation of process and product parts of the method fragment and, second, the implementation of fragments' storage, retrieval, and construction. Even if all ME approaches investigate storing methods fragments in the "method base" or "method repository" [5, 6, 14], this information is relevant as all the other implementation parts are founded on this one. Hence, our *tool/implementation* attribute takes the following values: product storage and manipulation, process operating, retrieval, and construction.

### 2.3 Subject View

This view answers the « What » question. This means that we will develop facets concerning the internal structure and formalisation of the fragment.

An observation of the literature guided us to define three possible *levels* in which we may consider fragments: intentional, structural, operational levels. The intentional level allows defining the context of use and/or reuse of fragments. The structural level determines the fragment structure and the kind of structural links between the fragment elements: specialization, composition and references. The operational level deals with operating part of fragment (allowing its implementation during development project).

The method fragment could be also characterized with relation to its main elements. Depending on the dominant element, [15] identifies three key *perspectives* for fragments description: process focussed, product focussed, and producer focussed.

Another important aspect of the fragment is the *recursion*. The concept of granularity is used in several approaches to allows the possibility to compose a fragment with others fragments. For instance, a fragment may be an entire method that can be decomposed in other less complex fragments (which, in turn, may also be decomposed in other more simple fragments).

A fragment may also be defined at different *abstraction levels*. We consider the following levels: meta-meta-model, meta-model, model [16].

[17] explores the notion of the fragment *formalism* that can be either conceptual when fragments are expressed with descriptions and specifications of methodology parts, or technical when there is an implementation of operational parts with tools.

### 2.4 Process View

The process view considers different ways of method fragments conception and usage. The attributes of this view aim at describing the main ME activities dealing with fragments (method decomposition, fragment selection, new method construction, and so on).

First, the methods are decomposed into methods fragments which are stored in method base (or repository). Thus, we define the facet "*decomposition principle*" which deals with different ways to decompose methods into fragments. This principle predefines the fragments description used for their identification during project fulfilment.

Once the methods are decomposed and stored in the base, they could be used in the projects. On the first step, the engineer must find in the method base the fragments that better match the project specificities. On this basis, we identify two facets: retrieval/selection principle and matching with situation. The *retrieval/selection principle* defines steps to carry out for identifying an appropriate fragment. In ME, all approaches are situational, which means they take into account the specific project situation by different manners. This aspect is considered within the *matching with situation* attribute.

The next step is to build a new method from the selected fragments. Based on [18], we distinguish the following main manners to use fragments for *constructing a new method* according to project specificities: assembly, extension, and reduction. By assembly, separate fragments are grouped with regard to the studied specific project to form a unique method [19]. By applying extension, a basic method is transformed into a new one by addition of new fragments [19]. By reduction, some fragments are removed from the basic method in order to transform it to match the engineer's needs [7]. In the real world projects, with time and resource constraints, where is a need for constructing methods dynamically depending on the project specificity and adapting it during its realization if project characteristics change. This property implies the agility of methods. Recently, the agility was discussed with regards to methods of IS development [20]. However, agility in ME approaches is not widely spread and is only suggested in recent works. To consider this kind of construction, we introduce the fourth value for the given attribute "*agile construction*" having a Boolean value.

## 3   Framework Application

Several types of fragments have emerged in the literature. The most known of these different kind of representation are method fragments, method chunks, component, OPF fragment, and method services [2]. Before applying our comparison framework to these fragments (sub-section 3.2), we give their brief overview in the following sub-section.

### 3.1   Overview of Existing Method Fragments

In order to succeed in creating good methodologies that best suit given situations, fragments representation and cataloguing are very important activities. In particular, they have to be represented in a uniform way that includes all the necessary information that may influence their retrieval, integration or assembling. The five above-mentioned method fragments are presented in the Figure 2. and quickly described below.

*Method fragments* (cf. Figure 2.A) [5, 21] are standardised building blocks based on a coherent part of method. A fragment is either a Product or a Process fragment and is stored on a method base from which they can be retrieved to construct a new method following assembly rules [17].

The latest description of a *method chunk* [2] describes it as a way to capture more of the situational aspects in ME and to appropriately support the retrieval process. A chunk [6] based method aims at associating the reusable components to their description in order to facilitate component research and extraction according to the user's needs (cf. Figure 2.B).

For [2], *method components* developed in [7, 22, 23] allow to view methods as constituted by exchangeable and reusable components. Each component consists of descriptions for process (rules and recommendations), notations (semantic, syntactic and symbolic rules for documentation), and concepts (cf. Figure 2.C). [23] introduces the notion of method rationale which is the systematic treatment of the arguments and reasons behind a particular method [22].

In the OPEN Process Framework (*OPF*) [8], the *fragment* is generated from an element in a prescribed underpinning meta-model [2]. This meta-model (cf. Figure 2.D) has been upgraded with the availability of the international standard ISO/IEC 24744 [24].

SO2M (Service Oriented Meta-Method) [9] develops a new kind of fragment called offers a repository with a large variety of method fragments, called *method services* together with, and a service composition process. During composition, the process guides developer's choices; it selects method services and delivers a method fragment that achieves a developer's requirement. The SO2M meta-model is based on three main principles: service orientation, task ontology for reuse of knowledge on development problems and dynamic construction of method services for generating tailored methods (cf. Figure 2.E).

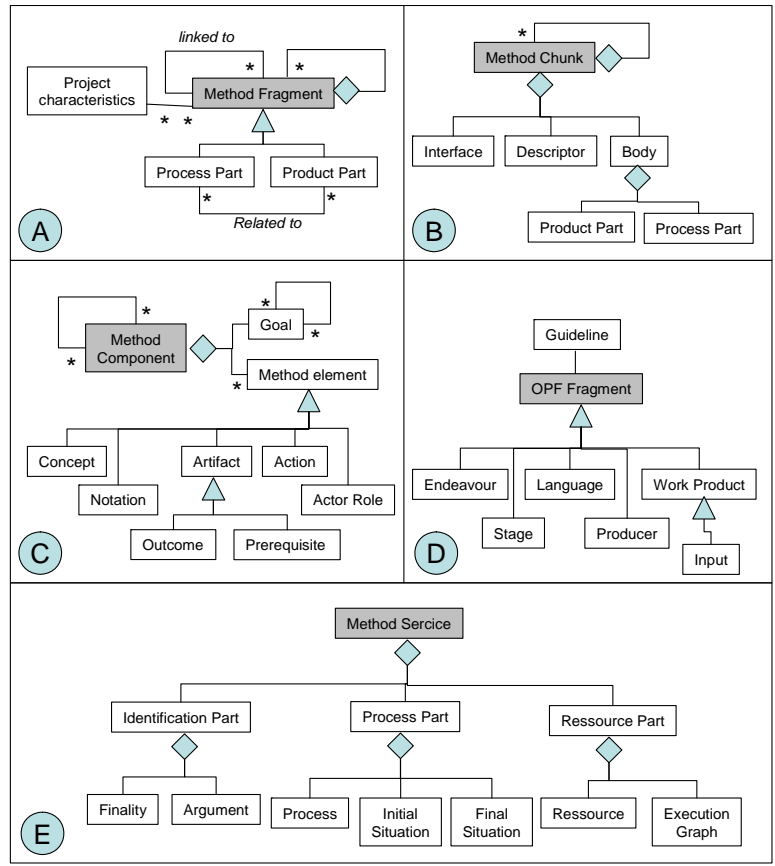

**Fig. 2.** Meta models parts of the reviewed fragment types.

### 3.2 Comparative Analysis within Framework

The table (cf. Appendix) presents a comparative analysis of the five selected fragments. This table is explained in this sub-section, attribute by attribute.

**Interoperability.** All fragments provide an internal interoperability, i.e. with fragments in the same method base. OPF fragments [14] can deal with an external interoperability in the same environment by using the object serialisation. Due to this serialisation, it can not be applied on different platforms. Method services [9] grant a fully external interoperability with a decentralised interoperable service oriented approach.

**Interactivity with user.** In most ME approaches, the creation, retrieval, composition, and application of fragments is done manually. Some efforts have been done with method fragments and method services to provide tools to assist the different users. However, most of their aspects are still done manually.

**Covered way.** All fragments help to construct methods that partly cover Seligmann definition of a method [13]. Indeed, each constructed method answers to a particular paradigm ('way of thinking') and has two different parts, namely the "product" ('way of modelling') and the "process" ('way of working'). However, not a single one of them is able to meet all the tool requirements ('way of supporting').

**Tools/implementation.** All considered ME approaches provide a tool for storing method fragments in a database. Method chunks also allow a more efficient retrieval of stored knowledge with the Method Chunk Repository [25]. Two other approaches go further in tool supporting. In addition to the fragments selection and retrieving, the first one (method fragments with a tool called Decamerone) contains the product part elements [5] and the second one (method services of SO2M approach) uses resource descriptions and execution graphs for implementing resource part [9]. However, the method service is viewed as a "black box" without any explanation on how it is developed. The OPF fragment authors develop an implemented product part within an "Eclipse" tool [26].

**Level.** The intentional level is present in all fragments excepted the method fragment one. The chunk's intentional level contains an interface (situational and intentional aspects) and a descriptor (set of criteria to locate the best engineering situation) [27]. For method component, the intentional level includes goal's identification. The OPF fragment is selected by its goal. The method service's identification part defines the purpose of the service: the finality (the problem that the method service solves) and the argument (advantages and drawbacks of using the method service). All fragments have a structural level. The operational part at the level of meta-model is included only in method service.

**Perspective.** The method fragments are defined as either process part or product part [5, 17], whereas all the other fragments include both the interrelated parts in their definition. The third perspective (producer) is addressed in only two blocks: component and OPF fragment [2]. In [5], roles of people are included as a property of the method fragment.

**Recursion.** Even if nearly all ME approaches insist on the different layers of fragment granularity (a fragment may be either a method part or a complete method [5,17]), only the method chunks can be described as completely recursive. A chunk is based on the decomposition of the method process model into reusable guidelines [28], which means that all chunks may be formally decomposed in other complete chunks. The other types of fragments are not formally defined to deal with process decomposition.

**Abstraction level.** All fragments are defined at the level of meta-models (cf. Figure 1.). The method service includes a meta meta model level because this approach suggests a ontology used for describing product model [9]. The OPF fragment contains also an endeavour, which is an instance of model and corresponds to a schema of development method [14].

**Formalism.** Chunks and components use conceptual formalisms, when the OPF fragments and method services support technical presentation. The method fragment contains both conceptual and technical representations [17].

**Decomposition Principle.** The decomposition principle is quite different following the fragment type. Method fragment uses a tree decomposition to link all coherent method parts. Chunks are obtained by intentional decomposition of methods

[19]. The OPF fragment is a "clabject", which is a result of both instantiation and inheritance [14]. Components are decomposed by goals [7]. The method service approach does not specify this attribute value.

**Retrieval/Selection Principle.** The retrieval and selection of a method fragment are made by different types of queries. Chunks are selected with the application of similarity measures of their descriptors and interfaces. This helps to evaluate the degree of matching between them and the requirements [19]. On the same way, the method service selection is made by a comparison of the requirements (expressed by intentions) with the service intentional descriptors by ontologies, which allow comparing the semantic similarity [9]. Differently, OPF fragments, stored on a 'work product tool', are selected with queries on their endeavour [14]. Method fragments are selected by application of request on the goal [21].

**Matching with situation.** Approaches don't match the situation with the same techniques. The method fragment definition consists in encouraging a global analysis of the project while basing itself on contingency criteria. Projects and situations are characterized by means of factors associated with the methods. The chunk approach includes projects requirements expressed as a *requirements map* [19], which is used to test the similarity between requirements and existing fragments. In component containing its "rational", the matching is performed by goal analysis [7]. The Method service approach uses an identification part that defines the purpose of the service. The matching is thus done by using goal, actor, process, and product ontologies [9].

**Construction technique.** The method fragments are assembled for creating a new method. The chunk approach uses assembly (allowing overlapping between different chunks) and extension. In addition to the assembling and extending, the component approach suggests method reduction. The method service construction is based on a composition process that supports the aggregation of services in sequence or in parallel [9]. In the OPF approach, a new method is constructed by dynamic instantiation of fragments during the project. Hence, the OPF approach suggests an agile construction of methods.

### 3.3. Drawbacks of Existing Method Fragments

The framework analysis allows identifying the following main drawbacks of existing method fragments. (i) The way of supporting method fragments is not sufficiently managed by ME approaches to produce new method with tool support. (ii) The ME approaches themselves are not enough automated. They limit their tool support to a description language, a method fragment repository, and retrieval facilities. (iii) Moreover, the handle of abstraction levels in fragments is not complete in all ME approaches. Fragments work at different abstraction level and the whole complexity of exchanged data is not addressed and causes a restriction of exchanges between them. (iv) Despite standardisation efforts of the ME community, there is no unified description language of a method fragment and interoperability issues between the various fragments method databases are not handled.

## 4 Improvement of the Method Service Concept

Our proposal to solve these problems is to carry on the approach proposed by C. Rolland in [10], i.e. to consider the method fragment as a service.

### 4.1 Proposed Solution

To develop the concept of method service, we use the Service Oriented Architecture (SOA) [11] and the method chunk definition.

Indeed, the SOA applied to ME needs may solve the limitation of existing ME approaches. The adaptation of the SOA to the ME – the Method Oriented Architecture (MOA) – defines a method services registry where a list of available method services is organised. This provides access to decentralised method service providers for ME engineers and developers and interoperable method services.

Moreover, according to the MOA, each method service has to be considered as a standalone component, which should be retrieved and selected dynamically. Each granularity of method service can be then viewed and executed as a method. To be compliant to this requirement, we based our method services on the method chunk for two reasons : (i) the intentionality (decomposition and retrieving). We decompose the methods into method services according to an intentional principle. We use the descriptor and interface of the method chunk in order to describe our method services intentional part. This part will be used for retrieving and selecting method services from the registry (ii) the recursion. Chunks use an intentional decomposition, which means that they are using the composition principle with a description of their intention (objective the engineer will reach if he uses it). To decompose a main intention into simpler ones allows a decomposition of a method into chunks logically related to each other and always described on the same way. This recursive description of the chunks will allow us to implement our services according to the MOA principles of process composition.

Satisfying the MOA requirements, our implementation of method services has to deal with the identified drawbacks by applying our comparison framework. To overcome them, a method service has to deal with four keys technical issues [29]: complexity, interoperability, composition, and interactivity. These issues can be addressed by the application of standards used in service-oriented approaches. Table 1. shows, for each issue, the suggested standards and their usage objectives.

The usage orientation is emphasised by this solution in several directions:

− Adoption of an open and distributed architecture to design, to distribute, and to execute method chunks.

− Enrichment of the semantic descriptor of method chunks with their corresponding software module, called method service.

− Adoption of standards widely used coming from web services technologies to implement method services.

**Table. 1.** Standards used for resolving the technical issues.

| Issue | Standard | Objective |
|---|---|---|
| exchanged data complexity | XMI – XML Metadata Interchange [30] | external data exchange on all levels |
| | MOF – Meta Object Facility [16] | modeling levels handling |
| interoperability | SOAP – Simple Object Access Protocol [31] | method services communication |
| | WSDL – Web Services Description Language [32] | method services descriptions |
| | UDDI – Universal Description Discovery and Integration [33] | service registry |
| | XMI | standardisation of exchanged products |
| | MOF | standardisation of exchanged products |
| composition | BPEL – Business Process Execution Language [34] | method services operational parts composition |
| interactive web services | WSRP – Web Services for Remote Portlets [35] | method services user interface handling |

### 4.2 Method Service Structure

The method service structure combines a *descriptor part* with its *implementation part* as shown on Figure 3. The descriptor part aims at documenting, retrieving, composing, and invoking the related implementation part. The tool support is realised by the implementation part of a method service. Each granularity level of *method service* is executable and may be a composition of method services.

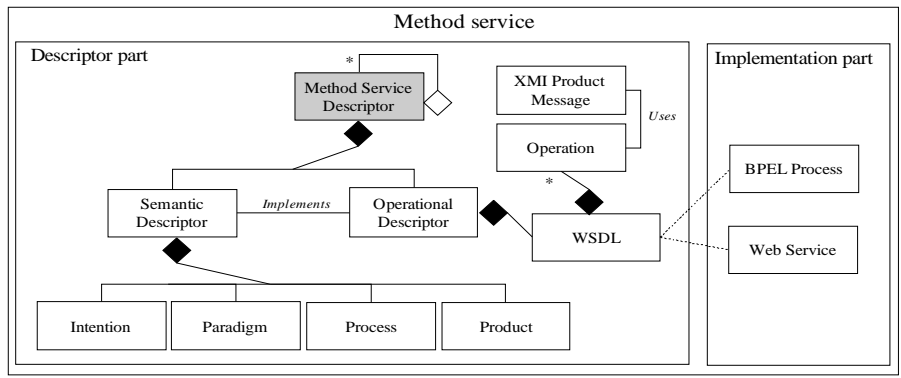

**Fig. 3.** Meta model of a method service fragment.

The *semantic descriptor* describes the chunk implemented by the method service. The main purpose of this descriptor is to document method services through four sub-parts: Intention, Paradigm, Process and Product. In the method chunk approach, the retrieval and composition of fragments are done by intentions. We propose to carry on this principle to base the retrieving and the composition of method services on the four sub-parts of the semantic descriptor.

The *intention* defines the intentions of the method service use and the context in which it can be reused. The *paradigm* describes the fragment's way of thinking. The

*process* is the description of activities executed on input products. The *product* is the meta-model description of input and output product models of the method service.

The operationalisation of method services is performed by an *operational descriptor* and an *implementation part*. It implements the process described in the semantic part by a *web service* or a composition of web services (*BPEL process*) exposed by a WSDL descriptor. The implemented web service is a tool providing the way of supporting method services. The product dimension is implemented by meta-models compliant to Meta Object Facility (MOF) standard and XMI schema standard.

The *WSDL* is the operational descriptor of a method service. It contains the definitions of each performed *operation* including their inputs, and outputs messages (*XMI product message*).

### 4.3 Method Oriented Architecture

As indicated above, the MOA (proposed in [10]) is an adaptation of the SOA to ME. Figure 4 shows the three actors and their interactions in the MOA: the *method provider*, the *method registry,* and the *method client*.

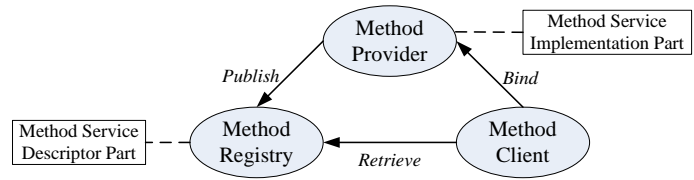

**Fig. 4.** Method Oriented Architecture.

The *method provider* creates method services and publishes their descriptors on the *method registry*. The *method client* retrieves method services from the registry, using retrieval facilities built upon semantic descriptors. Operational descriptors (WDSL) are used by method clients to invoke the *implementation part* of method services from their provider.

This implementation of the fragment process is either an atomic web service or a composition of web services realised by a BPEL process.

The MOA usage can be sketched according to two use cases:
– developers using CASE tools to invoke remote method services.
– method engineers using CAME tools to define new methods with method services composition facilities.

This MOA provides an open and decentralised access to method services for method client tools built on a Software as a Service (SaaS) architecture [36].

### 4.4 Method Service Characterization according to the Framework

A method service reuses the method chunks characteristics. Indeed, its semantic descriptor is inspired by method chunk descriptor. However, it includes also the way of supporting by its operational descriptor and its implementation part. We have defined our method service based on the comparison framework:

Objective view {*Interoperability* = "external in different environments"; *Interactivity with user* = "automated"}

Usage view {*Covered ways* = "thinking, modeling, working, supporting"; *Tools / implementation* = "storage, manipulation, operating, retrieval, construction"}

Subject view {*Level* = "intentional, structural, operational"; *Perspective* = "Process focussed, Product focussed"; *Recursion* = yes; *Abstraction level* = "meta-meta-model, meta-model, model, schema"; *Formalism* = "conceptual, technical"}

Process view {*Decomposition principle* = "by intentions"; *Retrieval/selection principle* = "request by paradigms, intentions, processes, products"; *Matching with situation* = "not specified"; *Construction technique* = "agile"}

We may observe with this definition that we have tried to overcome the drawbacks identified in the section 3.3. First of all, this fragment is a "real" method fragment as it covers the four parts of Seligmann definition [13] by developing the full support of the method service. The interoperability issue is ensured by the adoption of widely used standards, coming from the web service and from the meta data exchange technologies. Creation, retrieval, composition, and application of method fragments are automated in our MOA based approach. The intentional decomposition principle gives a recursive view and the fragment is viewed as a service. Finally, our suggestion allows an agile construction of situational methods.

### 4.5 Basic Application of the Method Service approach

The following figures illustrate our approach with the description and application of a method service called *Objectify* (Fig 5 and Fig. 6). This service implements the process of making out an object out of a relationship (known as objectification, reification, or nesting) [37].

Fig. 5 shows the semantic descriptor of this method service. The product part shows the input and the output class diagram parts whereas the process part shows the operations which has to be executed on the input product to reach the method service intention and obtain the output product.

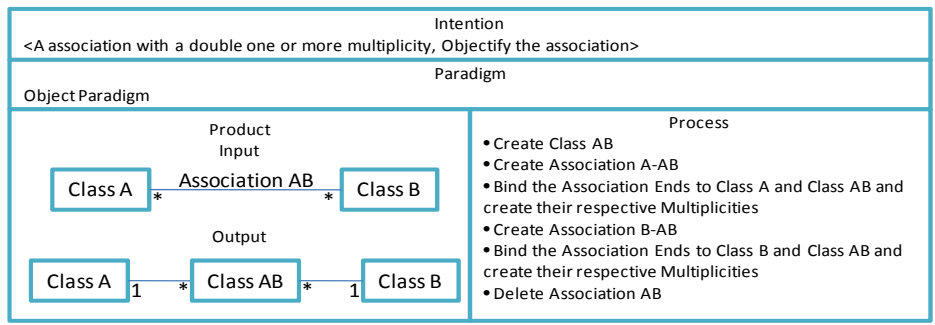

**Fig. 5.** *Objectify* Method Service Semantic Descriptor.

Fig. 6 shows a part of the method service implementation. We focus on the invocation of the web service implementation. There is other processes that have to be taken into account to implement this approach, as the search and retrieval of the descriptor (WSDL), but we thought that this one will be enough here to give a relevant example to illustrate this work.

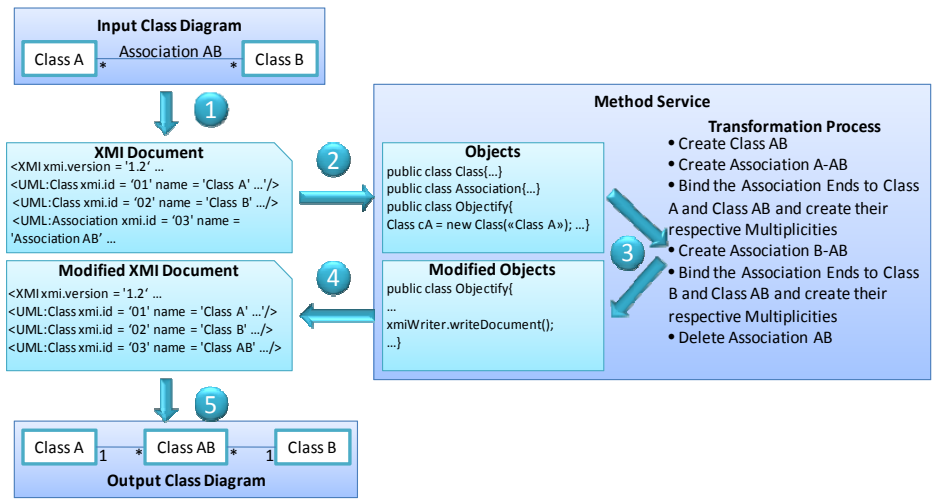

**Fig. 6.** *Objectify* Method Service Application Example.

As mentioned above, in table 1, we choose to use XMI standard for data exchange. Consequently, the input class diagram needs to be represented using XMI to produce a XMIDocument (step 1) which will be understandable by a method service. This one is the implementation of a method service process part and has to be applied on the XMIDocument. At this point, several implementation solutions are possible. For instance, the method service directly modify the XML code of the XMIDocument, either by an algorithm application, a modification of the DOM (Document Object Model) tree [38] or by an XSLT transformation [39], or it may be instantiated to manipulate objects. In our illustration, we choose this last solution because it induces a more easy transformation (step 2) and instantiate the XMIDocument according to the UML Meta Model (MOF compliant) [40]. Then, we manipulate the created objects by a simple algorithm in order to perform the chunk process part (step 3). Once modified, the instance of the input XMIDocument is used to generate the output XMIDocument (step 4) which represents the transformed class diagram (step 5).

## 5   Discussion: Toward a Unique Concept of Fragment

Different method fragments and their correlations represent a main purpose of ME science. An attempt to find a unique concept was made during the panel of the ME conference [2]. In this section, we present our point of view on this problem and discuss the possibility to lead to a unique vision of fragment with regard to the suggested definition of method service.

The creation of a unique concept will be confronted to several challenges to solve. The definition of a method by [13] decomposes a method in four ways, which have to be addressed by the unique concept of fragment. Furthermore, in a general way this unique fragment will have to cover all the concepts contained in actual fragments. Afterwards, in a practical view, the four technical issues enounced in section 4.1 (complexity, interoperability, composition, and interactive web services) have to be considered. Some advantages could be retrieved from a unique concept like the standardisation of method fragments providing an interoperability of solutions, encouraging the share and use of fragments.

Nevertheless, covering all aspects of method fragments in a unique fragment is a difficult task. Therefore, we propose to define some essential aspects required for a unique fragment representation. For the fragment purpose, five aspects have to be considered: intentionality, reusability, interoperability, interactivity, and implementation.

Our proposal of the method service improvement addresses most challenges of a unique fragment concept. The four ways of a method and technical issues are considered, but the covering all existing method fragments aspects is not provided. Therefore, intentionality and reusability objectives are not yet completely implemented. The implementation of our semantic descriptor and its associate platform will solve these two problems.

## 6 Conclusion

In this paper, our contribution is double: we define a comparison framework in order to identify the drawbacks of existing method fragments and propose an improvement of the method service concept to solve them.

The suggested framework allows a comparison structured in four views and the following purposes: (i) to have an overview of existing method fragments, (ii) to define drawbacks of existing method fragments, and (iii) to analyse the possibility to converge on a unique fragment concept.

Based on this framework analysis, we propose to improve the method service concept in order to:

− overcome the following drawbacks of existing method fragments with the application of service-oriented approaches standards: insufficient consideration of complexity, lacks of interoperability, and lacks of interactivity;

− encourage the usage of fragments with: the application of widely used standards, the providing of a tool support, and the adoption of a MOA providing an open and distributed architecture.

The current implementation of our approach allows method engineers to create method services. For now, we do not integrate the corresponding user interface with method services (back office services). These services may be used to modify existing methods or create new ones with BPEL processes. A limitation of our work is the implementation of the composition principle as we can only implement assembly composition without overlapping. This principle is a very big technical issue on which we are currently working.

Our future works include implementing the semantic part of method services and defining a way for characterising the specific project situation. Our aim is to build both the CASE tool based on SaaS for supporting new methods (created by the application of ME approaches) and the CAME tool for method engineers for composing method services using the semantic descriptors.

## References

1. Rolland, C.: L'ingénierie des méthodes : une visite guidée *(in French :* Method Engineering: A Guided Visit), e-TI - la revue électronique des technologies d'information, 1, http://www.revue-eti.netdocument.php?id=726 (2005)
2. Agerfalk, P., Brinkkemper, S., Gonzales-Perez, C., Henderson-Sellers, B., Karlsson, F., Kelly, S., Ralyté, J.: Modularization Constructs in Method Engineering: Towards Common Ground?, Panel of ME 07, Springer, Geneva, Switzerland, (2007)
3. Henderson-Sellers, B., Gonzalez-Perez, C., Ralyté, J.: Situational Method Engineering: Fragments or Chunks?, proceedings of CAiSE'07 Forum, Trondheim, Norway, (2007)
4. Aharoni, A., Reinhartz-Berger, I. : Representation of method Fragments, a comparative study, in proceedings ME 07, Springer, Geneva, Switzerland, (2007)
5. Brinkkemper, S.: Method Engineering: engineering of information systems development method and tools, Information and Software Technology, 38(7), (1996)
6. Rolland, C., Plihon, V., Ralyté, J.: Specifying the reuse context of scenario method chunks, in the proceedings of the international conference. CAiSE'98, Pise, (1998)
7. Wistrand, K., Karlsson, F.: Method components: Rationale revealed, in proceedings of CAISE 04, Springer-Verlag. Riga, Latvia, (2004)
8. Henderson-Sellers, B.: Process meta-modelling and process construction: examples using the OPF. Ann. Software Engineering, 14(1-4), (2002)
9. Guzélian, G., Cauvet, C.: SO2M : Towards a Service-Oriented Approach for Method Engineering, in: the 2007 World Congress in Computer Science, Computer Engineering and Applied Computing, in the proceedings of the international conference IKE'07, Las Vegas, Nevada, USA, (2007)
10. Rolland, C.: Method Engineering : Achievements,Trends & Challenges, In the keynote presentations of ME'07, (2007)
11. W3C: Web Services Architecture (WSA), http://www.w3c.org/TR/2004/NOTE-ws-arch-20040211/, (2004)
12. Rolland, C., Ben Achour, C., Cauvet, C., Ralyte, J., Sutcliffe, A., Maiden, N.M., Jarke, M., Haumer, P., Pohl, K., Dubois, E. and Heymans, P.: A Proposal for a Scenario Classification Framework, Requirements Engineering Journal (1998)
13. Seligmann, P.S., Wijers, G .M., Sol, H.G.: Analysing the structure of IS methodologies, an alternative approach, Proceedings of the 1st Dutch conference on Information Systems, Amersfoort, The Netherlands, (1989)
14. Gonzales-Perez, C.: Supporting Situational Method Engineering with ISO/IEC 24744 and the Work Product Tool Approach. Proceedings of the International IFIP WG8.1 Conference ME 07, Springer, Geneva, Switzerland, (2007)
15. Henderson-Sellers, B.: SPI – A role for Method Engineering, Proceedings of the 32nd EUROMICRO, SEAA'06, (2006)
16. Object Management Group (OMG): Meta Object Facility (MOF)v2.0, http://www.omg.org/spec/MOF/2.0/, (2006)

17. Brinkkemper, S., Saeki, M., Harmsen, A.F.: A method engineering Language for the description of systems development methods, in proceedings of the conference CAISE 01. Springer Verlag. Interlaken, Switzerland, (2001)
18. Nehan, Y.-R., Deneckère, R.: Component-based Situational Methods: A framework for understanding SME, in IFIP, Volume 244, Situational Method Engineering: Fundamentals and Experiences, Switzerland, (2007)
19. Ralyté, J., Deneckere, R., Rolland, C.: Towards a Generic Model for Situational Method Engineering, in proceedings of the conference CAISE'03, Springer Verlag, Velden, Austria, (2003)
20. Abrahamsson, P., Salo, O., Ronkainen, J., Warsta, J.: Agile Software Development Methods: Review and Analysis, VTT Publication 478 (2002)
21. Harmsen, A.F., Brinkkemper, J.N., Oei, J.L.H.: Situational Method Engineering for information Systems Project Approaches Int. IFIP WG8. 1 Conference in CRIS series: "Methods and associated Tools for the Information Systems Life Cycle" (A-55), North Holland (Pub.), (1994)
22. Agerfalk, P.J.: Information systems actability: Understanding Information Techology as a Tool for Business Action and Communication. Doctoral dissertation. Dept. of Computer and Information Science, Linköping University, (2003)
23. Karlsson, F.: Method Configuration: Method and Computerized Tool Support. Doctoral dissertation. Dept of Computer and Information Science. Linköping University. (2005)
24. International Standards Organization / International Electrotechnical Commission: Software Engineering. Metamodel for development Methodologies, ISO/IEC 24744, Geneva, (2007)
25. Jeusfeld, M., Backlund, P., Ralyté, J.: Classifying Interoperability Problems for a Method Chunk Repository. I-ESA'07, Funchal, Portugal, (2007)
26. Firesmith, D.: Method Engineering Using OPFRO, European SEPG, Netherlands (2006)
27. Mirbel, I. and Ralyte, J. -- Situational method engineering : combining assembly-based and roadmap-driven approaches -- Requirement Engineering Journal, 11(1), (2006)
28. Ralyte, J., Rolland, C.: An Approach for Method Reengineering. Conference on The Entity-Relationship Approach Yokohama, Japan (2001)
29. Souveyet, C., Iacovelli, A.: Method as a Service (MaaS), Submitted to the conference RCIS'08, (2008)
30. Object Management Group (OMG): XML Metadata Interchange (XMI) v2.1, http://www.omg.org/technology/documents/formal/xmi.htm, (2005)
31. Gudgin, M., Hadley, M., Mendelsohn, N., Moreau, J.J., Nielsen, H.F., Karmarkar, A., Lafon, Y.: SOAP V.1.2, in W3C Recommendations, http://www.w3.org/TR/soap/, (2007).
32. Christensen, E., Curbera, F., Meredith, G., Weerawarana, S.: Web Services Description Language (WSDL) 1.1, in W3C Notes, http://www.w3.org/TR/wsdl, (2001)
33. OASIS: UDDI Version 3.0.2, http://www.oasis-open.org/committees/uddi-spec/doc/spec/v3/uddi-v3.0.2-20041019.htm, (2004)
34. OASIS: Web Services Business Process Execution Language Version 2.0, http://docs.oasis-open.org/wsbpel/2.0/wsbpel-v2.0.pdf, (2007)
35. OASIS: Web Services for Remote Portlets Specification 1.0, http://www.oasis-open.org/committees/download.php/3343/oasis-200304-wsrp-specification-1.0.pdf (2003)
36. Papazoglou, M.P.: Service-Oriented Computing: Concepts, Characteristics and Directions, in the Proceedings of the Fourth International Conference on Web Information Systems Engineering (WISE 2003), (2003).
37. Halpin, T. : Objectification, Proceedings of the Tenth International Workshop on Exploring Modeling Methods in Systems Analysis and Design (EMMSAD'05), Porto, Portugal, 13-14 June (2005)
38. W3C: Document Object Model (DOM) : http://www.w3.org/DOM/ (2005)
39. W3C: XSL Transformations (XSLT) : http://www.w3.org/TR/xslt (1999)
40. Unified Modeling Language (UML) : http://www.uml.org/

**Appendix: Comparative Analysis of Five Selected Fragments**

| View | Attributes | Values domain | Fragment | Chunk | Component | OPF Fragment | Method Service |
|---|---|---|---|---|---|---|---|
| Objective | Interoperability | internal, external in the same environment, external in different environments | internal | internal | internal | external in the same environment | external in different environments |
| Usage | Interactivity | manual, assisted, automated | assisted | manual | manual | manual | assisted |
| | Covered way | thinking, modeling, working, supporting | thinking, modeling, working | thinking, modeling, working | thinking, modeling, working | thinking, modeling, working | thinking, modeling, working |
| | Tools / implementation | product storage, manipulation, process operating, retrieval, construction | storage, manipulation, retrieval, construction (Decamerone) | storage, retrieval (method chunk repository) | Not specified | Not specified | storage, retrieval, construction |
| Subject | Level | intentional, structural, operational | structural | intentional, structural | intentional, structural | intentional, structural | intentional, structural, operational |
| | Perspective | Process focused, Product focused, producer focused | Process and Product focused | Process and Product focused | Process, Product, and Producer focused | Process, Product, and Producer focused | Process and Product focused |
| | Recursion | booleen | no | yes | no | no | no |
| | Abstraction level | meta-meta-model, meta-model, model | meta-model, model | meta-model, model | meta-model, model | meta-model, schema | meta-meta-model, meta-model, model |
| | Formalism | conceptual, technical | conceptual | conceptual | conceptual | technical | technical |
| Process | Decomposition principle | | | by intentions | by goal | inheritance, instantiation | Not specified |
| | Retrieval/selection principle | | request | similarity measure | request by goal | request by goal | semantic similarity |
| | Matching with situation | | project characterisation | requirements map | | by goal and actor | by goal, actor, process, and product ontologies |
| | Construction technique | assembly, extension, reduction, agile | assembly | assembly, extension | assembly, extension, reduction | agile | assembly without overlapping |